\documentclass[showpacs,amsmath,amssymb,aps,showkeys,floatfix,prd,a4paper]{revtex4}

\usepackage[dvips]{graphicx}
\usepackage{dcolumn}
\usepackage{bm}
\usepackage{epsfig}
\usepackage{amsfonts}
\usepackage{amssymb,amscd}

\def\lsim{\raise0.3ex\hbox{$<$\kern-0.75em\raise-1.1ex\hbox{$\sim$}}}
\def\gsim{\raise0.3ex\hbox{$>$\kern-0.75em\raise-1.1ex\hbox{$\sim$}}}

\def\pom{{I\!\!P}}

\def\beq{\begin{equation}}
\def\eeq{\end{equation}}
\def\bea{\begin{eqnarray}}
\def\eea{\end{eqnarray}}
\def\bq{\begin{quote}}
\def\eq{\end{quote}}

\def\gappeq{\mathrel{\rlap {\raise.5ex\hbox{$>$}}
{\lower.5ex\hbox{$\sim$}}}}

\def\lappeq{\mathrel{\rlap{\raise.5ex\hbox{$<$}}
{\lower.5ex\hbox{$\sim$}}}}

\def\Toprel#1\over#2{\mathrel{\mathop{#2}\limits^{#1}}}

\def\pom{{I\!\!P}}

\begin{document}


\title{Diffractive quarkonium photoproduction in $pp$ and $pA$ collisions at the LHC: Predictions of the Resolved Pomeron model  for the Run 2 energies}

\author{V.~P. Gon\c{c}alves, L. S. Martins and B. D. Moreira}
\affiliation{High and Medium Energy Group, \\
Instituto de F\'{\i}sica e Matem\'atica, Universidade Federal de Pelotas\\
Caixa Postal 354, CEP 96010-900, Pelotas, RS, Brazil}

\date{\today}

\begin{abstract}
The inclusive diffractive quarkonium photoproduction in $pp$ and $pA$ collisions is investigated considering the Resolved Pomeron Model to describe the diffractive interaction.  We estimate the rapidity and transverse momentum distributions for the $J/\Psi$, $\Psi(2S)$ and $\Upsilon$ photoproduction in hadronic collisions at the LHC and present our estimate for the total cross sections at the Run 2 energies.  A comparison with the predictions associated to the exclusive production also is presented. Our results indicate that the inclusive diffractive production is a factor $\gtrsim 10$ smaller than the exclusive one in the kinematical range probed by the LHC.
 
\end{abstract}
\keywords{Ultraperipheral Heavy Ion Collisions, Vector Meson Production, QCD dynamics}
\pacs{12.38.-t; 13.60.Le; 13.60.Hb}

\maketitle

\section{Introduction}
\label{intro}

The treatment of diffractive processes have attracted much attention as a way of amplifying the physics programme at hadronic colliders, including searching for New Physics (For a recent review see, e.g. Ref. \cite{review_forward}). The investigation of these reactions at high energies gives important information about  the structure of hadrons and their interaction mechanisms. In particular, hard diffractive processes allow the study of the interplay of small- and large-distance dynamics within Quantum Chromodynamics (QCD). The diffractive processes can be classified as {\it inclusive} or {\it exclusive} events (See e.g. \cite{forshaw}). In exclusive events, empty regions  in pseudo-rapidity, called rapidity gaps, separate the intact very forward hadron from the central massive object. Exclusivity means that nothing else is produced except the leading hadrons and the central object. The inclusive diffractive processes also exhibit rapidity gaps. However,  they contain soft particles accompanying the production of a hard diffractive object, with the rapidity gaps becoming,  in general, smaller than in the exclusive case. 

During the last years, the study of exclusive processes in photon -- induced interactions at hadronic colliders \cite{upc} became a reality \cite{cdf,star,phenix,alice,alice2,lhcb,lhcb2,lhcb3,lhcbconf} and new data associated to the Run 2 of the LHC are expected to be released soon. In particular, there is the expectation that the experimental data for the exclusive vector meson photoproduction in $pp/pA/AA$ collisions will allows to constrain the main aspects of the treatment of 
the QCD dynamics at high energies and large nuclei (See e.g. Refs. \cite{bert,vicmag,brunoall,brunorun2,vicdiego}). As demonstrated in Ref. \cite{brunorun2}, the color dipole model description of the exclusive vector meson photoproduction  allows to describe the  Run 1 data, as well as the preliminary data on  $pp$ collisions at $\sqrt{s} = 13$ TeV, if the non -- linear effects are taken into account in the QCD dynamics. In this model, the diffractive interaction is described in terms of a Pomeron exchange, which is represented at lowest order by the two -- gluon color singlet state. At higher orders, the description of the color singlet interaction is directly associated to the modelling of the QCD dynamics \cite{hdqcd}.

\begin{figure}[t]
\includegraphics[scale=0.35]{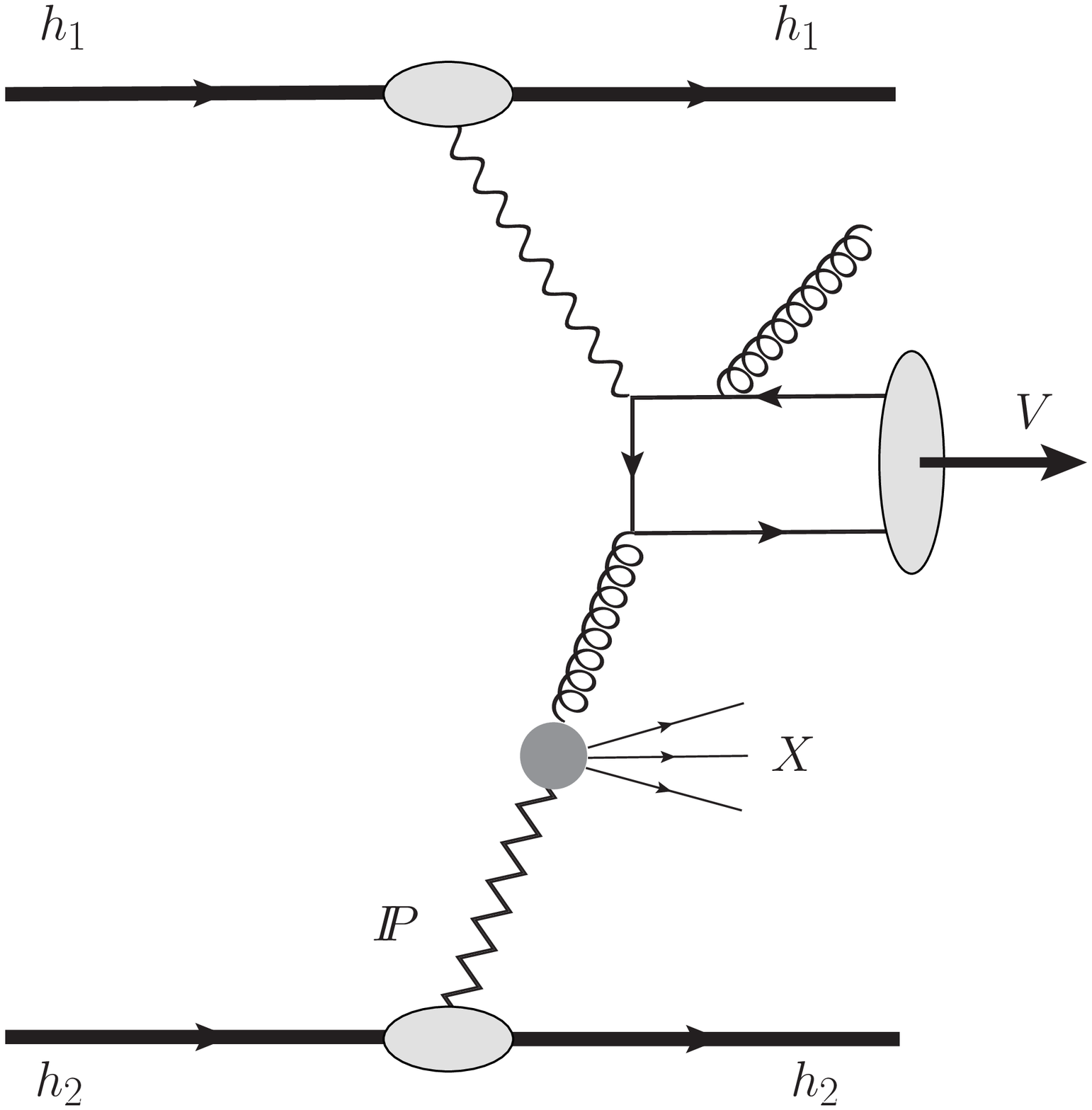}
\includegraphics[scale=0.35]{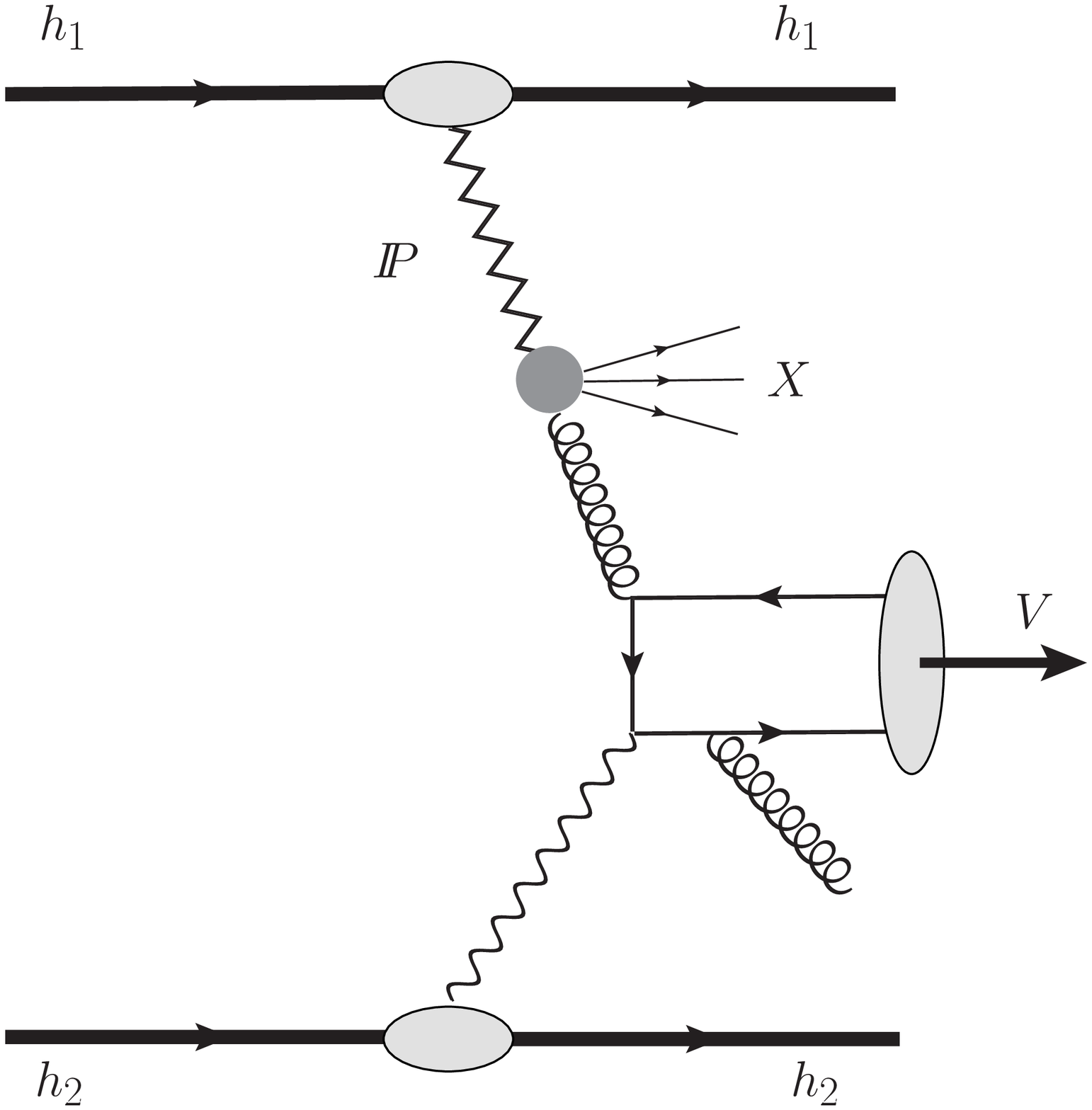} 
\caption{Schematic view of  typical diagrams for the inclusive diffractive quarkonium photoproduction in hadronic collisions considering the Resolved Pomeron model for the diffractive interaction and the NRQCD formalism for the quarkonium production.}
\label{fig1}
\end{figure}

As pointed above, in addition to the exclusive production, a given final state also can be produced in an inclusive diffractive interaction. As observed in $ep$ collisions at HERA and hadronic  collisions at Tevatron and LHC, this contribution can be important in some regions of the phase space (See e.g. Ref. \cite{scho}). In particular, the recent ZEUS data for the diffractive photoproduction of isolated photons \cite{zeus} indicate that both contributions are important for the description of the process. The inclusive diffractive production of a given final state is  in general calculated 
 assuming the validity of the diffractive factorization formalism and that Pomeron has a partonic structure. The basic idea is that  the hard scattering resolves the quark and gluon content in the Pomeron \cite{IS} and  it can be obtained analysing the experimental data from diffractive deep inelastic scattering (DDIS) at HERA, providing us with the diffractive distributions of singlet quarks and gluons in the Pomeron \cite{H1diff}. This model is usually denoted Resolved Pomeron model. During the last years, this model have been applied to estimate the single and double diffractive production of different final states 
\cite{MMM1,antoni,cristiano2,vicmurilo}. Our goal in this paper is to extend this formalism for the quarkonium photoproduction and estimate, by the first time, the inclusive diffractive contribution for the   photoproduction of  $J/\Psi$, $\Psi(2S)$ and $\Upsilon$ in $pp$ and $pA$ collisions at the LHC. In the Resolved Pomeron model, this process is represented  by the diagrams presented in Fig. \ref{fig1}. Similarly to the exclusive production, this process also will be characterized   by two rapidity gaps and  two intact hadrons in the final state. However, in addition to the vector meson,  the remnants of the Pomeron also are expected to be present in the final state in inclusive interactions.  Moreover, as we will show in the next Section, a gluon also is expected to be present in the final state. The presence of the remnants and the gluon should increase the number of tracks in the detector. Therefore, in principle, the inclusive and exclusive contributions could be separated using the exclusivity criteria. If this separation is feasible, is also allows to study in more detail the modelling of the inclusive diffractive processes and the  description of the quarkonium production. On the other hand, due to the large pile up of events in each bunching crossing expected for the Run 2, it is not clear if the separation of the inclusive diffractive events will be possible by measuring the rapidity gaps and counting the number of tracks in the final state. In principle, the only possibility to  detect double diffractive events, as those associated to the inclusive and exclusive diffractive vector meson photoproduction, is by tagging the intact hadrons in the final state. It implies the key element to measure diffractive events at the LHC will be tagging  the  forward scattered incoming hadrons \cite{detectors}. In this case, it is important to determine the background to exclusive events associated to the inclusive contribution. All these aspects motivate the analysis that will be performed in what follows.
 
This paper is organized as follows. In the next Section we will discuss the photon - induced interactions at the LHC and present a brief review of the Resolved Pomeron Model for treatment of diffractive interactions as well the NRQCD formalism for the quarkonium production. In Section \ref{res} we present our predictions for the rapidity and transverse momentum distributions and total cross sections considering $pp$ and $pA$ collisions at the Run 2 energies of the LHC. Moreover, our predictions will be compared with those for the exclusive production. Finally, in Section \ref{conc} we summarize our main conclusions.

\section{Resolved Pomeron Model description of the Diffractive Quarkonium Photoproduction} 

In this Section we will present a brief review of the main concepts needed to describe the photon -- induced interactions in hadronic collisions and the 
formalism used in our calculations of the quarkonium photoproduction.
The basic idea in  photon-induced processes is that an ultra relativistic charged hadron (proton or nucleus) 
gives rise to strong electromagnetic fields, such that the photon stemming from the electromagnetic field of one of the two colliding hadrons can 
interact with one photon of the other hadron (photon - photon process) or can interact directly with the other hadron (photon - hadron process) \cite{upc}. 
In these processes the total cross section  can be factorized in terms of the equivalent flux of photons into the hadron projectile and the photon-photon 
or photon-target  cross section. In this paper we focus on the inclusive diffractive quarkonium production in photon -- hadron interactions at hadronic collisions. 
A schematic view of the process considered in this paper is presented in Fig. \ref{fig1}. As in the exclusive production, the final state for the inclusive process will be characterized by two rapidity gaps in the final state, one associated to the photon exchange and another to the Pomeron one. The cross section for the  quarkonium photoproduction  will be given by,
\begin{equation}
   \sigma(h_1+h_2 \rightarrow h_1 \otimes H + X \otimes h_2) =  \int d \omega \frac{dN}{d\omega}|_{h_1}
   \, \sigma_{\gamma h_2 \rightarrow H X \otimes h_2}\left(W_{\gamma h_2}  \right) + \int d \omega \frac{dN}{d\omega}|_{h_2}
   \, \sigma_{\gamma h_1 \rightarrow H X \otimes h_1}\left(W_{\gamma h_1}  \right)\,  \; , 
\label{eq:sigma_pp}
\end{equation}
where $\otimes$ represents the presence of a rapidity gap in the final state, $\omega$ is the photon energy in the center-of-mass frame (c.m.s.), $\frac{dN}{d\omega}|_{h_i}$ is the equivalent photon flux for the hadron $h_i$, $W_{\gamma h}$ is the c.m.s. photon-hadron energy given by $W_{\gamma h}=[2\,\omega\sqrt{s}]^{1/2}$, where
$\sqrt{s}$ is  the c.m.s energy of the
hadron-hadron system. 
Considering the requirement that  photoproduction
is not accompanied by hadronic interaction (ultra-peripheral
collision) an analytic approximation for the equivalent photon flux of a nuclei can be calculated, which is given by \cite{upc}
\begin{eqnarray}
\frac{dN}{d\omega}|_{A}= \frac{2\,Z^2\alpha_{em}}{\pi\,\omega}\, \left[\bar{\eta}\,K_0\,(\bar{\eta})\, K_1\,(\bar{\eta}) - \frac{\bar{\eta}^2}{2}\,{\cal{U}}(\bar{\eta}) \right]\,
\label{fluxint}
\end{eqnarray}
where   $K_0(\eta)$ and  $K_1(\eta)$ are the
modified Bessel functions, $\bar{\eta}=\omega\,(R_{h_1}+R_{h_2})/\gamma_L$ and  ${\cal{U}}(\bar{\eta}) = K_1^2\,(\bar{\eta})-  K_0^2\,(\bar{\eta})$.
 On the other hand, for   proton-proton interactions, we assume that the  photon spectrum is given by  \cite{Dress},
\begin{eqnarray}
\frac{dN}{d\omega}|_{p} =  \frac{\alpha_{\mathrm{em}}}{2 \pi\, \omega} \left[ 1 + \left(1 -
\frac{2\,\omega}{\sqrt{S_{NN}}}\right)^2 \right] 
\left( \ln{\Omega} - \frac{11}{6} + \frac{3}{\Omega}  - \frac{3}{2 \,\Omega^2} + \frac{1}{3 \,\Omega^3} \right) \,,
\label{eq:photon_spectrum}
\end{eqnarray}
with the notation $\Omega = 1 + [\,(0.71 \,\mathrm{GeV}^2)/Q_{\mathrm{min}}^2\,]$ and $Q_{\mathrm{min}}^2= \omega^2/[\,\gamma_L^2 \,(1-2\,\omega /\sqrt{s})\,] \approx (\omega/
\gamma_L)^2$, where $\gamma_L$ is the Lorentz boost  of a single beam. This expression  is derived considering the Weizs\"{a}cker-Williams method of virtual photons and using an elastic proton form factor (For more details see Refs. \cite{Dress,Kniehl}). 
Equation (\ref{eq:sigma_pp}) takes into account the fact that the incoming hadrons can act as both target and photon emitter.  In our calculations of  the inclusive diffractive quarkonium photoproduction 
in hadronic collisions we will assume that the rapidity gap survival probability $S^2$ (associated to probability of the scattered proton not to dissociate 
due to  secondary interactions) is equal to the  unity. The inclusion of these  absorption effects in $\gamma h$ interactions is still a subject of intense 
debate \cite{Schafer_ups,frankfurt_ups,Martin}.

The main input in our calculations is the inclusive diffractive quarkonium photoproduction  cross section, $\sigma_{\gamma + h \rightarrow H + X \otimes h}$. In order to estimate this quantity we need to specify the model that will be used to describe the quarkonium photoproduction and the diffractive interaction. In the last decades, a number of theoretical approaches have been proposed for the calculation of the heavy quarkonium production, as for instance,  the Non Relativistic QCD (NRQCD)  approach, the fragmentation approach, the color singlet model (CSM), the color evaporation model and the $k_T$-factorization approach  (For a review see, e.g., Ref. \cite{review_nrqcd}).
In the specific case of the  non - diffractive quarkonium photoproduction, the description of the experimental data for this process is a challenge to the distinct approaches, as verified in Ref. \cite{h1} and discussed in detail in Refs. \cite{review_nrqcd,bk}. Keeping in mind that the underlying mechanism governing heavy quarkonium production is still subject of intense debate, in what follows we will describe this process in terms of the NRQCD formalim \cite{nrqcd}. In this formalism,  the cross section for the production of a heavy quarkonium state $H$ factorizes as  $\sigma (ab \rightarrow H+X)=\sum_n \sigma(ab \rightarrow Q\bar{Q}[n] + X) \langle {\cal{O}}^H[n]\rangle$, where the coefficients $\sigma(ab \rightarrow Q\bar{Q}[n] + X)$ are perturbatively calculated short distance cross sections for the production of the heavy quark pair $Q\bar{Q}$ in an intermediate Fock state $n$, which does not have to be color neutral.  The $\langle {\cal{O}}^H[n]\rangle$
are nonperturbative long distance matrix elements, which describe the transition of the intermediate $Q\bar{Q}$ in the physical state $H$ via soft gluon radiation. Currently, these elements have to be extracted in a global fit to quarkonium data as performed, for instance, in Ref. \cite{bk}. 
  In this paper we will extend the NRQCD approach for the diffractive quarkonium photoproduction. Moreover, as in Ref. \cite{h1}, we will estimate the cross section for $z < 1$, which suppress the contribution of the $2 \rightarrow 1$ subprocess, associated to the $ \gamma + g \rightarrow  H$ channel. As a consequence,
 the total cross section for the $\gamma h \rightarrow H + X \otimes h$ process can be expressed at leading order  as follows (See e.g. \cite{ko})
\begin{eqnarray}
\sigma (\gamma + h \rightarrow H + X \otimes h) = \int dz dp_{T}^2 \frac{xg^D(x,Q^2)}{z(1-z)} 
 \frac{d\sigma}{d \hat{t} }(\gamma + g \rightarrow H + g ) \label{sigmagamp}
\end{eqnarray}
where $z \equiv (p_H.p)/(p_{\gamma}.p)$, with $p_H$, $p$ and $p_{\gamma}$ being the four momentum of the quarkonium, hadron and photon, respectively. In the hadron rest frame, $z$ can be interpreted as the fraction of the photon energy carried away by the quarkonium. Moreover, $p_{T}$ is the magnitude of the quarkonium three-momentum normal to the beam axis and $g^D$ is the diffractive gluon distribution, which will be modelled using the Resolved Pomeron Model, to be discussed in more detail below.  Consequently, the partonic differential cross section $d\sigma/d \hat{t}$  is given by \cite{ko} 
\begin{eqnarray}
\frac{d\sigma}{d \hat{t} }(\gamma + g \rightarrow H + g)
=  \sum_n \,\, \langle {\cal{O}}^H[n]\rangle \, \cdot \, \frac{d\sigma}{d \hat{t} }(\gamma + g \rightarrow c\bar{c}[n] + g) \,\,,
\end{eqnarray}
which takes into account the color -- singlet and color -- octet contributions for the quarkonium production. The color -- singlet contributions can be expressed as follows \cite{ko}
\begin{eqnarray}
\frac{d\sigma}{d \hat{t} }(\gamma + g \rightarrow H + g)|_{singlet} = 
   \frac{1}{16\pi\hat{s}^{2}}
 \overline{\sum} \left|   
 {\cal M} 
 \right|^{2} 
 \left( 
 \hat{s},\hat{t}
 \right)              
\end{eqnarray}
where
\begin{eqnarray}
 \overline{\sum} \left|   
 {\cal M} 
 \left( 
 \gamma g \rightarrow H g 
 \right)
 \right|^{2}
 =
 {\cal N}_{1} 
 \frac{\hat{s}^{2} \left( \hat{s} -4M_{Q}^{2} \right)^{2} 
 + \hat{t}^{2} \left( \hat{t} -4M_{Q}^{2} \right)^{2} 
 +\hat{u}^{2} \left( \hat{u} -4M_{Q}^{2} \right)^{2}}
 {\left( \hat{s} - 4M_{Q}^{2} \right)^{2} 
 \left( \hat{t} - 4M_{Q}^{2} \right)^{2}
 \left( \hat{u} - 4M_{Q}^{2} \right)^{2}} \,\,.
\end{eqnarray}
The normalization factor ${\cal N}_{1}$ is given by 
\begin{eqnarray}
 {\cal N}_{1} = 
 \frac{32}{9} 
 \left( 4 \pi \alpha_{s} \right)^{2} 
 \left( 4 \pi \alpha \right) 
 e_{Q}^{2} M_{Q}^{3} 
 G_{1}\left( H \right) \,\,\, ,
\end{eqnarray}
where $e_Q$ and $m_Q$ are, respectively, the charge and mass of heavy quark constituent of the quarkonium. Moreover, the factor $G_1$ is directly related to the color singlet matrix element ${\langle H |  {\cal O}_{1}\left( ^{3}S_{1}  \right)  | H \rangle}$ as follows
\begin{eqnarray}
 G_{1} \left( H \right) = 
 \frac{\langle H |  {\cal O}_{1}\left( ^{3}S_{1}  \right)  | H \rangle}
 {M_{Q}^{2}} \,\,\, . 
\end{eqnarray}
On the other hand, the octet contribution is given by \cite{ko,klasen}
\begin{eqnarray}
\frac{d\sigma}{d \hat{t} }(\gamma + g \rightarrow H + g)|_{octet} = \frac{1}{16\pi \hat{s}^{2}} 
\left\{  \left( 4 \pi \alpha_{s} \right)^{2} 
 \left( 4 \pi \alpha \right) 
 e_{Q}^{2} \, 
\frac{\langle H | {\cal O}^{H}(^{2S+1}L_{J}^{[8]}) |H  \rangle}{(2J+1)M_{Q}} 
\times f (^{2S+1}L_{J}^{[8]})   
\right\}  
\end{eqnarray}
with
\begin{eqnarray}
f (^{2S+1}L_{J}^{[8]}) = 
\frac{3 \hat{s} \hat{u}}{2\hat{t}(\hat{s} + \hat{t})^{2} (\hat{t} + \hat{u})^{2} (\hat{u} + \hat{s})^{2}}   \left[  
\hat{s}^{4} + \hat{t}^{4} + \hat{u}^{4} + \left( 2M_{Q} \right)^{8}
 \right].
  \label{amp2_octeto}
\end{eqnarray}

Lets now discuss the modelling of the diffractive gluon distribution in the Resolved Pomeron Model \cite{IS}. In this model,  the diffractive parton distributions are expressed in terms of parton distributions in the \,{Pomeron} and a Regge parametrization of the flux factor describing the \,{Pomeron} emission by the hadron. The  parton distributions have evolution given by the DGLAP evolution equations and should be determined from events with a rapidity gap or a intact hadron. The diffractive gluon distribution, $g^D_{p} (x,Q^2)$, is defined as a convolution of the \,{Pomeron} flux emitted by the proton, $f^{p}_{\pom}(x_{\pom})$, and the gluon distribution in the \,{Pomeron}, $g_{\pom}(\beta, Q^2)$,  where $\beta$ is the momentum fraction carried by the partons inside the \,{Pomeron}. 
The \,{Pomeron} flux is given by
\begin{eqnarray}
f^{p}_{\pom}(x_{\pom})= \int_{t_{\rm min}}^{t_{\rm max}} dt \, f_{\pom/{p}}(x_{{\pom}}, t) = 
\int_{t_{\rm min}}^{t_{\rm max}} dt \, \frac{A_{\pom} \, e^{B_{\pom} t}}{x_{\pom}^{2\alpha_{\pom} (t)-1}}  \,\,,
\label{fluxpom:proton}
\end{eqnarray}
where $t_{\rm min}$, $t_{\rm max}$ are kinematic boundaries. The \,{Pomeron} flux factor is motivated by Regge theory, where the \,{Pomeron} trajectory is assumed to be linear, $\alpha_{\pom} (t)= \alpha_{\pom} (0) + \alpha_{\pom}^\prime t$, and the parameters $B_{\pom}$, $\alpha_{\pom}^\prime$ and their uncertainties are obtained from fits to H1 data  \cite{H1diff}. \,{ The slope of the Pomeron flux is $B_{\pom}=5.5^{-2.0}_{+0.7}$ GeV$^{-2}$, the Regge trajectory of the Pomeron is
$\alpha_{\mathbb P}(t)=\alpha_{\mathbb P}(0)+\alpha_{\mathbb P}'~t$ with $\alpha_{\mathbb P}(0)=1.111 \pm 0.007$ and $\alpha_{\mathbb P}'=0.06^{+0.19}_{-0.06}$ GeV$^{-2}$. The $t$ integration boundaries are $t_{\rm max}=-m_{p}^2x_{\pom}^2/(1\!-\!x_{\pom})$ ($m_{p}$ denotes the proton mass) and $t_{\rm min}=-1$ GeV$^2$.} Finally, the normalization factor $A_{\mathbb P}=1.7101$ is chosen such that $x_{\pom}\times\int_{t_{\rm{min}}}^{t_{\rm{max}}}dt~f_{\pom/{p}}(x_{\pom},t)=1$ at $x_{\pom} = 0.003$.
The diffractive gluon distribution of the proton is then given by
\begin{eqnarray}
{ g^D_{p}(x,Q^2)}=\int dx_{\pom}~d\beta ~\delta (x-x_{\pom}\beta)~f^{p}_{\pom}(x_{\pom})~g_{\pom}(\beta, Q^2)={ \int_x^1 \frac{dx_{\pom}}{x_{\pom}} f^{p}_{\pom}(x_{\pom}) ~g_{\pom}\left(\frac{x}{x_{\pom}}, Q^2\right)} \,\,.
\label{difgluon:proton}
\end{eqnarray}
In our analysis we use the diffractive gluon distribution obtained by the H1 Collaboration at DESY-HERA \cite{H1diff}.

\begin{figure}[t]

\includegraphics[scale=0.4]{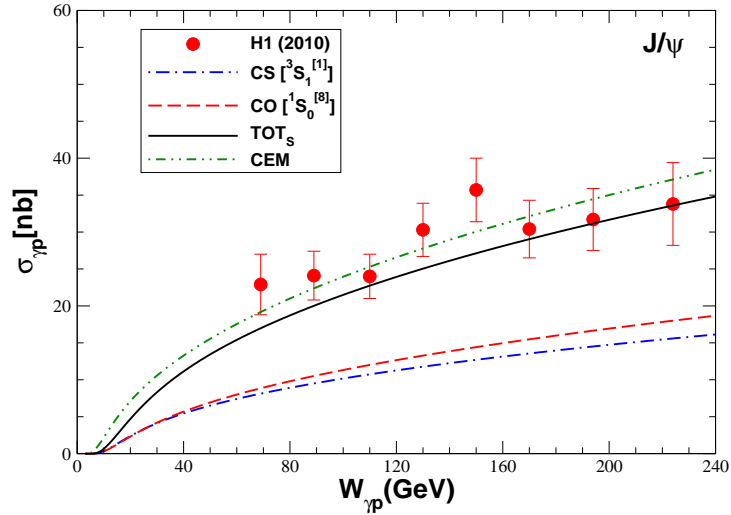}
 \caption{Predictions for the energy dependence of the cross section for the inelastic $J/\Psi$ photoproduction in inclusive $\gamma p$ interactions at HERA. Data from H1 Collaboration \cite{h1}.}
\label{fig0}
\end{figure}

\section{Results}
\label{res}
In this Section we will present our predictions for the rapidity and transverse momentum distributions for the inclusive diffractive $J/\Psi$, $\Psi(2S)$ and $\Upsilon$ photoproduction 
in $pp$ and $pPb$ collisions at the LHC energies. In the case of $pPb$ collisions, the cross sections will be dominated by $\gamma p$ interactions, due to the $Z^2$ enhancement present in the nuclear photon flux. As a consequence, the associated rapidity distributions will be asymmetric. We will assume $m_c = 1.5$ GeV and $m_b = 4.5$ GeV. Moreover, following Ref. \cite{elements}, we will consider that (in units of GeV$^3$):  
$\langle J/\Psi |  {\cal O}_{1}\left( ^{3}S^{[1]}_{1}  \right)  | J/\Psi \rangle = 1.2$, 
$\langle \Psi(2S) |  {\cal O}_{1}\left( ^{3}S^{[1]}_{1}  \right)  | \Psi(2S) \rangle = 0.76$,  
$\langle \Upsilon |  {\cal O}_{1}\left( ^{3}S^{[1]}_{1}  \right)  | \Upsilon \rangle = 10.9$,
$\langle J/\Psi |  {\cal O}_{1}\left( ^{1}S^{[8]}_{0}  \right)  | J/\Psi \rangle = 0.018$, 
$\langle \Psi(2S) |  {\cal O}_{1}\left( ^{1}S^{[8]}_{0}  \right)  | \Psi(2S) \rangle = 0.008$, and   
$\langle \Upsilon |  {\cal O}_{1}\left( ^{1}S^{[8]}_{0}  \right)  | \Upsilon \rangle = 0.0121$.  The calculations for the inclusive diffractive production  will be performed using the fit A for the diffractive gluon distribution \cite{H1diff}. We checked that the predictions are increased by $\approx 10 \%$ if the  fit B is used as input. Following Ref. \cite{h1} we will  integrate  the fraction of the photon energy carried away by the quarkonium in the  range  $0.3  \lesssim z \lesssim 0.9$ and we will take the minimum value of the transverse momentum of the quarkonium as being $p_{T,min} = 1$ GeV. As demonstrated in Ref. \cite{vicmairon}, the predictions are not strongly dependent on the  inferior limit of integration $z_{min}$. Finally, for comparison, we also will present the predictions associated to the exclusive quarkonium photoproduction derived in Ref.  \cite{brunorun2} using the dipole approach and the bCGC model for the dipole - proton cross section (See Ref. \cite{brunorun2} for details).

Before to present our predictions for the diffractive case, lets estimate the cross section for the non -- diffractive production in $\gamma p$ interactions at HERA using the NRQCD formalism and compare with the H1 data \cite{h1}. Such comparison is an important check of our calculations. In Fig. \ref{fig0} we present separately the singlet and color octet contributions, as well as the sum of both contributions. We have assumed $Q^2 = p_T^2 + m_c^2$ and the CTEQ6LO parametrization \cite{cteq6} for the inclusive gluon distribution. We have that the singlet and octet contributions are similar in the energy range considered and that the experimental data only are described if both contributions are taken into account. Consequently, in order to get realistic predictions of the inclusive diffractive production we will include both in our calculations. As discussed before, the modelling of the quarkonium production is still a theme of intense debate. In order to get an estimate of the theoretical uncertainty associated to the choice of the model used to describe the quarkonium production, in Fig. \ref{fig0} we also present the predictions obtained using the Color Evaporation Model (CEM), which was proposed many years ago \cite{velhos_cem}, extensively used in the literature \cite{cem_gregores,criscem} and recently improved \cite{ramona}. One have that this model also is able to describe the data, as expected from the analysis performed in Ref. \cite{cem_gregores}. In comparison with the predictions obtained using the NRQCD formalism,  its predictions are almost 15 \% larger at large energies. We have checked that a similar increasing also is observed in the predictions for the rapidity distributions and total cross sections obtained using the NRQCD formalism that will present below.

\begin{figure}[t]
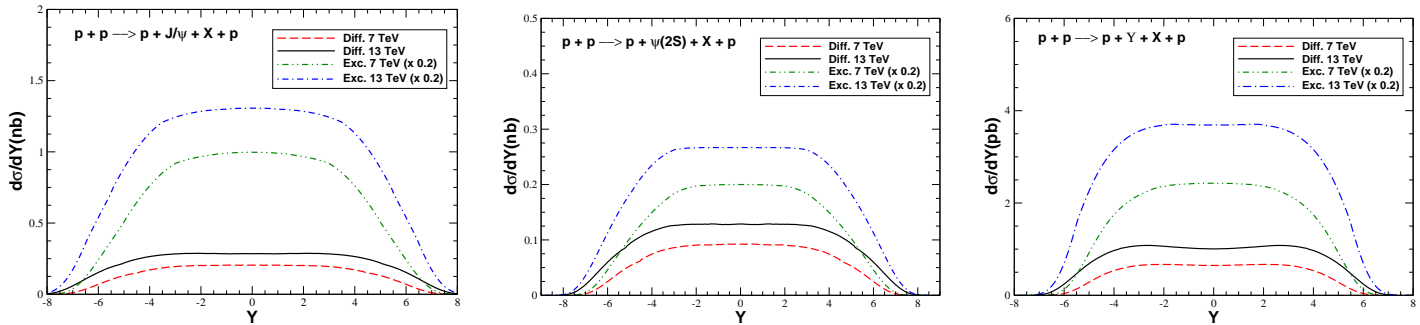

\begin{tabular}{ccccc}
\includegraphics[scale=0.25]{jpsi_diff_pp.eps} & \,\,\,\,\, & 
\includegraphics[scale=0.25]{psi2S_diff_pp.eps} & \,\,\,\,\, &
\includegraphics[scale=0.25]{upsilon_diff_pp.eps}
\end{tabular}
\caption{Rapidity distributions for the inclusive diffractive $J/\Psi$ (left panel), $\Psi(2S)$ (central panel) and $\Upsilon$ (right panel) photoproduction in $pp$  collisions at $\sqrt{s} = 7$ TeV (dashed lines) and 13 TeV (solid lines). The predictions associated to the exclusive quarkonium photoproduction  at $\sqrt{s} = 7$ TeV (dot - dot - dashed lines) and $\sqrt{s} = 13$ TeV (dot - dashed lines), multiplied by a factor 0.2, are also presented for comparison.}
\label{fig:rappp}
\end{figure}

 In Fig. \ref{fig:rappp} we present our predictions for the rapidity distributions 
 for the inclusive diffractive $J/\Psi$, $\Psi(2S)$ and $\Upsilon$ photoproduction in $pp$  collisions at $\sqrt{s} = 7$  and 13 TeV. As expected from Eq. (\ref{sigmagamp}) we have that the cross sections decrease with the mass of the vector mesons and increase with the energy. Moreover, we have symmetric rapidity distributions, which is directly associated to the fact that both the incident protons are sources of photons with the two terms in Eq. (\ref{eq:sigma_pp}) contributing equally at forward and backward rapidities, respectively. In comparison with the  predictions for the exclusive photoproduction, presented in Fig. \ref{fig:rappp} rescaled by a factor 0.2, we have that the rapidity distributions associated to inclusive diffractive production have a similar shape, but are smaller by a factor $\gtrsim 10$ at central rapidities, with the larger difference occuring in the case of the $J/\Psi$ production.	Such differences also are present in the predictions for the total cross sections shown in  Table \ref{tab1}.

\begin{widetext}
\begin{table}[t]
\centering
\begin{tabular}{|c|c|c|c|}\hline

                              &   $J/\psi$                        &     $\psi (2S)$                    &        $\Upsilon$                   \\ \hline
$pp$ ($\sqrt{s} = $ 7 TeV)    &    2.18 nb (49.21 nb)             &     0.94 nb  (9.83 nb)            &    6.30 pb (109.57 pb)                   \\ \hline
$pp$ ($\sqrt{s} = $ 13 TeV)   &    3.40 nb (72.21 nb)             &     1.47 nb (14.95 nb)              &    11.14 pb (189.50 pb)    \\ \hline
$pPb$ ($\sqrt{s} = $ 5 TeV)   &    1.63 $\mu$b     (47.45 $\mu$b) &   0.70 $\mu$b (8.67 $\mu$b)        &    2.92 nb (59.00 nb)                           \\ \hline
$pPb$ ($\sqrt{s} = $ 8.1 TeV) &    2.56  $\mu$b  (67.95 $\mu$b)   &  1.10 $\mu$b  (12.73 $\mu$b)       &    5.41 nb (104.69 nb)                         \\ \hline

\end{tabular} 

\caption{Total cross sections for the inclusive diffractive $J/\Psi$, $\Psi(2S)$ and $\Upsilon$ photoproduction in $pp$ and $pPb$ collisions at the Run 2 LHC energies. For comparison the predictions associated to the exclusive production  are also presented in parenthesis.}  
\label{tab1}
\end{table}
\end{widetext}

\begin{widetext}

\begin{figure}[t]
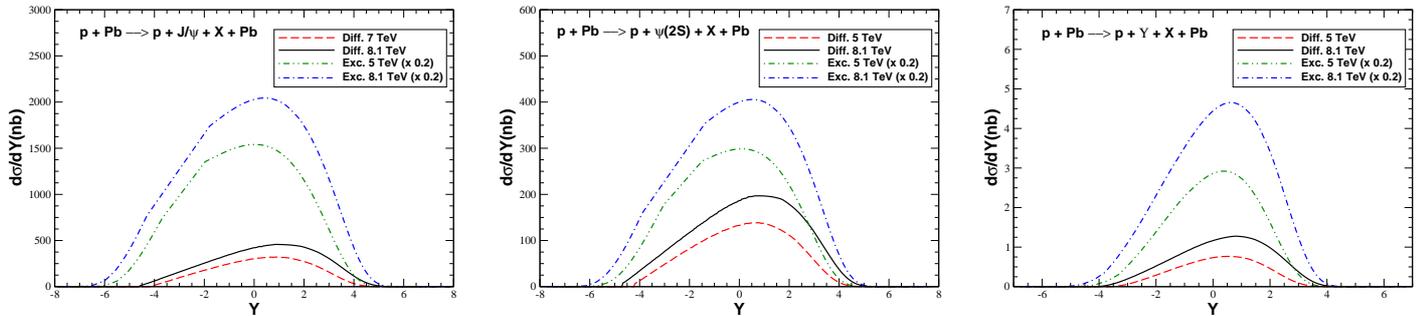

\begin{tabular}{ccccc}
\includegraphics[scale=0.25]{jpsi_diff_pA.eps} & \,\,\,\,\, & 
\includegraphics[scale=0.25]{psi2S_diff_pA.eps} & \,\,\,\,\, &
\includegraphics[scale=0.25]{upsilon_diff_pA.eps}
\end{tabular}
\caption{Rapidity distributions for the inclusive diffractive $J/\Psi$ (left panel), $\Psi(2S)$ (central panel) and $\Upsilon$ (right panel) photoproduction in $pPb$  collisions at $\sqrt{s} = 5$ TeV (dashed lines) and 8.1 TeV (solid lines). The predictions associated to the exclusive quarkonium photoproduction  at $\sqrt{s} =$ 5 TeV (dot - dot - dashed lines) and $\sqrt{s} =$ 8.1 TeV (dot - dashed lines),  multiplied by a factor 0.2, are also presented for comparison. }
\label{fig3}
\end{figure}

\end{widetext}

The predictions for the rapidity distributions in $pPb$ collisions at $\sqrt{s} = 5$ and 8.1 TeV are presented in Fig. \ref{fig3} and for the total cross sections in Table \ref{tab1}. As expected from the different photon fluxes for the proton and nuclei, we obtain asymmetric distributions. Due to the $Z^2$ enhancement on the nuclear photon flux, the total cross sections predicted for $pPb$ collisions are of the order of $\mu$b for the charmonium production.   As in the $pp$ case, the shape of the distributions for the inclusive and exclusive production are similar, with the inclusive diffractive predictions being smaller than the exclusive one by a factor $\gtrsim 10$ at $Y \approx 0$. Although the range of $\gamma p$ center - of - mass energies ($W_{\gamma p}$) that contribute in $pp$ and $pA$ collisions are distinct \cite{upc}, we have obtained that the differences between the diffractive and exclusive predictions are similar in both cases. This result is directly associated to the fact that both cross sections, although described in terms of distinct approaches, increase with the energy as $W_{\gamma p}^{\alpha_i}$ with a similar exponent for a given final state $i$.

\begin{widetext}

\begin{figure}[t]
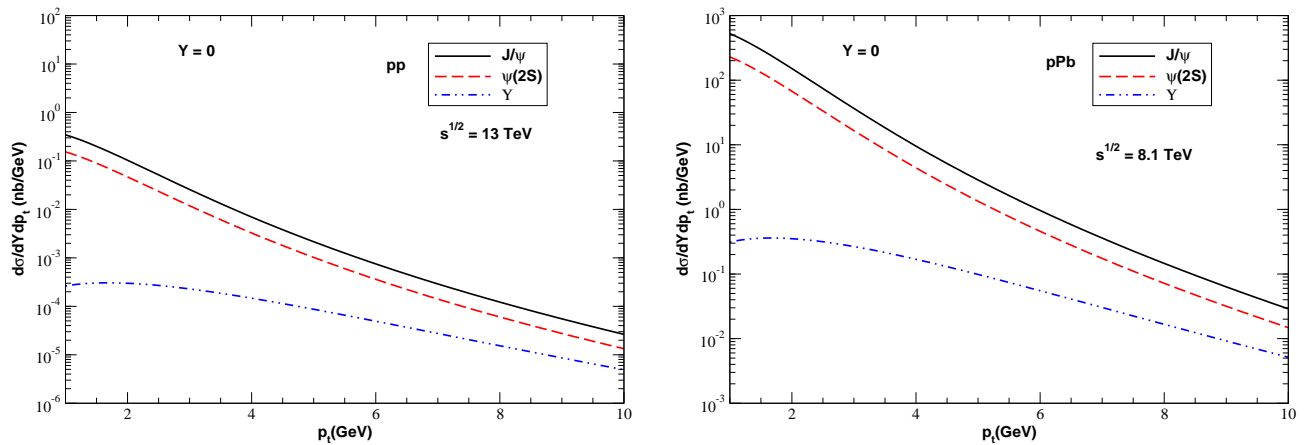

\begin{tabular}{ccc}
\includegraphics[scale=0.35]{pp_y0_13000_dist_pt.eps} & \,\,\,\,\, & 
\includegraphics[scale=0.35]{pA_y0_8100_dist_pt.eps}
\end{tabular}
\caption{Transverse momentum distributions for the inclusive diffractive $J/\Psi$, $\Psi(2S)$ and $\Upsilon$  photoproduction at central rapidities ($Y = 0$) in $pp$ collisions at $\sqrt{s} = 13$ TeV (left panel) and  $pPb$  collisions at $\sqrt{s} = 8.1$ TeV (right panel).}
\label{fig4}
\end{figure}

\end{widetext}

In Fig. \ref{fig4} we present our predictions for the transverse momentum distributions for the inclusive diffractive $J/\Psi$, $\Psi(2S)$ and $\Upsilon$  photoproduction at central rapidities ($Y = 0$) in $pp$ collisions at $\sqrt{s} = 13$ TeV (left panel) and  $pPb$  collisions at $\sqrt{s} = 8.1$ TeV (right panel). We have that the $p_T$ distributions for the $J/\Psi$ and $\Psi(2S)$ production are similar and differ only in  normalization. We predict a flatter distribution for the $\Upsilon$ production, which is directly associated to the larger quark mass for the bottom in comparison to the charm. Moreover, we predict that the transverse momentum distributions for  the diffractive quarkonium production decrease with $p_T$ following a power - law behavior $\propto 1/p_T^n$, where the effective power $n$ is energy dependent and is distinct for each  meson.  In contrast, in the exclusive production we have that the typical transverse momentum of the vector mesons in the final state is determined by the transferred momentum  in the Pomeron - proton vertex $\sqrt{|t|}$ (See e.g. Refs \cite{vicgustavo,vicdiego}). As the exclusive cross section has an $e^{ - \beta_V |t|}$ behavior, where $\beta_V$ is the slope parameter associated to the meson $V$,  the associated $p_T$ distribution of the vector mesons decreases exponentially at large transverse momentum. Therefore, it is expected that the production of quarkonia with a large $p_T$ should be dominated by the inclusive diffractive mechanism.

\section{Summary}
\label{conc}

During the last years, the experimental results from Tevatron, RHIC and LHC have demonstrated that the study of  hadronic physics using photon induced 
interactions in $pp/pA/AA$ colliders is feasible. In particular,  $\gamma h$ interactions at LHC  has been used to study the exclusive photoproduction of vector mesons, which is considered a probe of the QCD dynamics at high energies. This processes is characterized by two rapidity gaps and intact hadrons in the final state. In this paper we have estimated, by the first time, the inclusive diffractive quarkonium photoproduction using the Resolved Pomeron model to describe the diffractive interaction. This process generates a final state similar to the exclusive one. However,  additional tracks, associated to remnants of the Pomeron, are expected to  be present in the final state. Moreover, the size of the rapidity gaps are expected to be smaller in the inclusive diffractive case. We have estimated the rapidity and transverse momentum distributions for the diffractive $J/\Psi$, $\Psi(2S)$ and $\Upsilon$ photoproduction in $pp$ and $pPb$ collisions at the LHC. Our results indicate that the rapidity distributions for the exclusive and inclusive processes are similar, but differ in normalization. In particular, the inclusive diffractive predictions are a factor $> 10$ smaller than the exclusive one. However, as the transverse momentum distributions for the inclusive diffractive production has a power -- law behavior, which differ from the exponential behaviour present in the exclusive case, the inclusive diffractive mechanism become dominant for the production of vector mesons with a large - $p_T$.  Such aspect can be explorated in the future to separate the events associated to the inclusive diffractive production, which will allow to study in more detail the description of the diffractive interactions as well the mechanism of quarkonium production.


\section*{Acknowledgements}
VPG acknowledge useful discussions with M. M. Machado in the initial stages of this project.  This work was partially financed by the Brazilian funding agencies CAPES and CNPq.



\end{document}